\documentstyle[aps,prl,multicol]{revtex}

\begin {document}
\preprint{}

\title{A re-examination of the electronic structure of
$Bi_{2}Sr_{2}CaCu_{2}O_{8+\delta}$ and $Bi_{2}Sr_{2}Cu_{1}O_{6+\delta}$ -
An electron-like Fermi Surface and
the absence of flat bands at $E_F$.}
\author{Y. -D. Chuang$^1$, A. D. Gromko$^1$, D. S. Dessau$^1$, Y.
Aiura$^{2}$, Y.
Yamaguchi$^{2}$, K. Oka$^{2}$, A. J. Arko$^{3}$, J. Joyce$^{3}$, H.
Eisaki$^{4}$, S.I. Uchida$^{4}$, K. Nakamura$^{5}$, Yoichi Ando$^{5}$
}
\address{$^1$Department of Physics, University of Colorado, Boulder,
Colorado, 80309-0390}
\address{$^2$Electrotechnical Laboratory (ETL), 1-1-4 Umezono, Tsukuba,
Ibaraki 305, Japan}
\address{$^3$Los Alamos National Laboratory, Los Alamos, NM 87545}
\address{$^4$Department of Superconductivity, University of Tokyo,
7-3-1 Hongo, Bunkyo-ku, Tokyo 113, Japan}
\address{$^5$Central Research Institute of Electric Power Industry
(CRIEPI), 2-11-1 Iwato-Kita, Komae, Tokyo 201-8511, Japan}

%\date{Received \today}
\maketitle

%--------1-------------------3-------------------5-------------------7---------8

\begin{abstract}
We present a re-examination of the electronic structure and Fermi
Surface (FS) of Bi-Sr-Ca-Cu-O (BSCCO) as obtained from
angle-resolved photoemission experiments.  By applying a stricter set
of FS crossing criteria as well as by varying the incident photon
energy outside the usual range, we have found very different behavior from
that previously observed.  In particular we have found an electron-like FS
centered
around the $\Gamma$ point, and the flat bands at $E_{F}$ near the
$\bar{M}$ point of the zone are absent.  These results are robust
over a large range of dopings and from single to double layer samples.

\end{abstract}
%79.60.-i: Photoemission and photoelectron spectra
%71.30.+h: Metal-insulator transitions and other electronic
%transitions
\vspace*{-0.3 in}
\begin{multicols}{2}
%\narrowtext

Angle-Resolved Photoemission Spectroscopy (ARPES) has emerged as one
of the most
powerful tools for unearthing the electronic structure and  physics
of the high temperature superconductors (HTSC) and other correlated
electron systems
since it allows one to directly probe the
$\vec{k}$ space information of the electronic structure\cite{1}.
Major discoveries obtained from ARPES experiments on the HTSC's  have
included the
observation of flat bands (an extended van Hove singularity) at or very near
the Fermi level\cite{2,3,4,5}, a
superconducting gap with d-wave symmetry\cite{6,7},
and an anomalous pseudogap above $T_{c}$\cite{8,9,10}. Such
advances are all predicated on a thorough knowledge of the normal-state
FS topology of these superconductors, which has been almost
universally accepted to be a hole-like pocket centered around the
Brillouin zone corners (($\pi,\pi$) or X, Y points)\cite{11,12,13,14}.
In this Letter we argue that this most fundamental assumption of the
FS topology is incorrect or at least greatly oversimplified.
This finding should have a major impact on many previous studies and
theories, including all those concentrating on the flat bands at $E_F$,
the superconducting gap, and on the normal state pseudogap.

The measurements we report on here were taken on high quality
single crystals of the BiSrCaCuO (BSCCO) family of cuprate
superconductors.  These are nearly ideal materials for surface
sensitive ARPES studies because of the beautiful cleaved surfaces
that are obtainable.  Thus the ARPES information is usually interpreted as
being representative of the bulk physical properties of BSCCO.

To date ARPES data from BSCCO have been mostly limited to
a narrow photon energy range between $19$ and $25eV$.  This has been
largely a matter of convenience as well as the general expectation
that the physics observed at other photon energy ranges would be
essentially unchanged.  We have measured BSCCO at photon energies well
outside this range and have found very different physics from that
observed between $19$ and $25eV$.  Most of the data presented in this paper was
taken with a photon energy of $33eV$, at which we find qualitatively
different but much clearer behavior than that observed between $19$ and
$25eV$. Namely, the data shows with high clarity the existence
of an electron-pocket Fermi surface as well as the absence of flat
bands at $E_{F}$.

The experiments were mostly performed at the Synchrotron Radiation Center
in Wisconsin with a few backup experiments performed at the Stanford
Synchrotron Radiation Laboratory in
California.  At both labs we used VSW 50mm hemispherical energy analyzers
mounted on two-axis goniometers. The total experimental energy resolution was
about $50meV$ FWHM and the angular resolution was $\pm 1^{o}$.
All data shown in this paper was taken at or near 100K,
comfortably in the normal state. Most of the $Bi_{2}Sr_{2}CaCu_{2}O_{8+\delta}$
(Bi2212) samples for this study came from CRIEPI, with a portion
coming from ETL. The $Bi_{2}Sr_{2}Cu_{1}O_{6+\delta}$ (Bi2201) sample
studied came from the University of Tokyo.

Fig. 1(a) shows Ding et al's version of the generally accepted
hole-like FS topology of $Bi_{2}Sr_{2}CaCu_{2}O_{8+\delta}$
(Bi2212), from one of the most complete and heavily referenced
data sets available\cite{11}.
The thick lines represent the FS due to the main $CuO_{2}$ band.
The thin lines are FS replicas due to the superstructure modulation
with a $Q=(0.2\pi,0.2\pi)$. Support for such a topology is displayed in
Fig. 1(c) and
(d). Here we have Energy Distribution Curves (EDCs) from Ding taken
at hv=19 eV along the high symmetry cuts $\Gamma -\bar{M}$ and $\bar{M} -
X$, where $\Gamma = (0,0)$
and $\bar M = (\pi,0)$.  From these two panels, the authors claim there is
no main band FS
crossing along $\Gamma - \bar{M}$ but that there is one along $\bar{M} -
X$, as evidenced by the
relatively rapid loss of peak weight in this region.

Figure 1(e) presents a small portion of our new normal-state ARPES data at
$33eV$ obtained from an overdoped Bi2212 sample along the high
symmetry cut $\Gamma -\bar{M} - Z$ ($Z=(2\pi,0)$).
The energy scale for these spectra is the
same as in Ding's data in panels (c) and (d).  Our peaks are
sharper and better resolved than Ding's, showing the very high
quality of the data set.  Clear dispersion is observed with the
peak reaching $E_{F}$ near $.81\pi$ (red curve).  At this same k-value, the
peak rapidly loses weight,
indicating a FS crossing. The peak reappears in the second
zone and disperses back towards higher binding energy
(BE). Although only  a single cut, this data already indicates that
there is a fundamental difference between our data at the new photon
energy of $33eV$, and the previous data of Ding et al.  Our data indicates
main-band FS crossings along $\Gamma -\bar{M} - Z$, while they are not expected
according to the accepted hole-like FS topology.
In addition, the flat bands at $E_F$ at $\bar{M}$ (extended van-Hove
singularity)
observed near $20eV$\cite{2,4} have been replaced at $33eV$
by a strongly dispersive band which crosses $E_{F}$.

To analyze the data in more detail we have made $\vec k$-space plots of the
integrated spectral intensity $n(\vec k)$ as well as the weight right at
$E_{F}$ which we term $E_{F}(\vec k)$.  We first illustrate how $n(\vec k)$
and $E_{F}(\vec k)$
are expected to behave for a simple band-like state crossing the FS.
Fig. 2(a) shows the zero-temperature $n(\vec k)$ and $E_{F}(\vec k)$ for a
non-interacting
system. In this case there is only weight at
$E=E_{F}$ when $\vec k = \vec k_{F}$.  In Fig. 2(b) we introduce
interactions, which within the Fermi Liquid theory framework will reduce
both the weight of the delta function peak in $E_{f}(\vec k)$
and the step in $n(\vec k)$ to the value Z (quasiparticle
weight)\cite{16}.
Finite experimental energy and momentum resolution will broaden both
curves, as illustrated in Fig. 2(c). Finally, polarization
and matrix element effects will slowly alter $n(\vec k)$, as shown in Fig.
2(d).  It is clear from the above plots that at a true FS crossing the
following criteria should be obeyed: (1) $E_{F}(\vec k)$ should be
maximal. (2) $n(\vec k)$ should be at 50\% of it's maximal value, or
equivalently at the maximal gradient point. Also, at $E_{F}$ we expect (3)
the peak dispersion
to extrapolate to zero energy, and (4) the midpoint of the leading edge of
the spectrum to be at or
even beyond (on the unoccupied side of) $E_{F}$.

Panels (e)-(i) of figure 2 show real $n(\vec k)$ and $E_{F}(\vec k)$ plots
from our new data on the BSCCO family.  To our knowledge, this is the
first time that $n(\vec k)$ and $E_{F}(\vec k)$ plots have been
analyzed together, which turns out to be a powerful new tool to
obtain FS crossings.  To determine $n(\vec k)$ we integrated
the ARPES spectral weight from $-500meV$ to $+100meV$ so as to span the full
energy width of the peak, and for $E_{F}(\vec k)$ we
integrated over a $50meV$ wide window centered at $E_{F}$. All plots
were normalized so that the maximum weight along $\Gamma -\bar{M} - Z$
was set to $1$. According to criteria (1) and (2), a FS crossing
should occur when $n(\vec k)$ loses half of
its maximum value (excluding the background) and $E_{F}(\vec k)$
simultaneously
peaks. These points are indicated in the figure by the dashed green lines.
Panel (e) shows $n(\vec k)$ and $E_{F}(\vec k)$ obtained from the raw
data of Fig. 1 (e).  As expected, this way of analyzing the data
(criteria 1 and 2) gives identical FS
crossings as obtained by studying the peak dispersion (criteria 3 and
4).  Importantly, the drastic drop in $n(\vec k)$ at $\bar{M}$
to a value comparable to that at $\Gamma$ or $Z$ indicates that this
crossing must be due to
the main band and not due to the crossing of a weak superstructure
band.  Also, the drop in $n(\vec k)$ at
the $Z$ point to a level equivalent to that at the $\Gamma$ point can not
be explained
by photon polarization or orbital symmetry arguments\cite{17}.
Although not central to this paper, this observation
includes new physics which warrants much further experimental and
theoretical attention.

Panels (f) and (g) show data taken at hv=$22 eV$ from
a similar sample used to make the $33eV$ data of panel (e).  Our raw
$22eV$ data looks qualitatively similar to Ding's data of figure
1(b)\cite{11} or to other
previously published data\cite{2,4,15} including the presence of flat bands at
$E_F$ near $\bar M$. The data of panels (f) and (g) show
that the peak in $E_{F}(\vec k)$ and the $50\%$
point of $n(\vec k)$ do not coincide, but rather the peaks of each
coincide. This behavior is extremely
unusual.  Furthermore, the data of panel (f) is
highly asymmetric about
$\bar M$ (in the 2d approximation $\Gamma$ would be equivalent to $Z$
and $\bar M$ would be a real high symmetry point), in contrast to the
much more
symmetric behavior observed at $33eV$. This strange
behavior makes us question the nature of the states probed near $\bar
M$ at photon energies near $22$ eV.

The interesting behavior of the data at $33 eV$ calls for a complete mapping
of the FS topology at this new photon energy.  We have taken many cuts
over the Brillouin zone
on a slightly overdoped Bi2212 sample (different from that used in figures
(1) and (2)).  Within experimental resolution, each of the FS
crossing criteria gave
identical crossing locations for each cut.  The crossing
points indicated from these cuts are shown in
Fig. 3.  We have connected these points with thick
lines indicating the main FS (stronger ARPES peaks) and thin lines
indicating the superstructure-derived FS (ARPES peaks of about
$30\%$ the intensity of the main peaks).  The superstructure FS's are
seen to be replicas of the main FS's, but shifted by $\pm
0.2(\pi,\pi)$.  We note that the FS topology measured here is
closed around the $\Gamma$ point, i.e. it is electron-like.  This is
manifestly different from the ``accepted'' hole-like FS topology of
Bi2212.

Despite this discrepancy, there is data in the
literature that confirms our new FS topology.  Figure 3 shows an
overlay of our new FS with an $E_{F}$ intensity plot from an optimally
doped Bi2212
crystal at 33 eV measured by Saini et al\cite{12}. The $E_{F}$ intensity
is analogous to our $E_{F}(\vec k)$), i.e. the maximum
intensity locations should correspond to FS crossings.
The overlay shows a striking similarity between these
two independently obtained results.  However, instead of interpreting
their data as an electron-like FS, Saini et al argued that it was still
representative of a hole-like FS centered around $X$ and $Y$.  They
attributed the loss of weight at $\bar{M}$ to a pseudogap
in the spectral function, and further claimed that the high intensity cusps
orthogonal to the $\Gamma -Y$ direction were due to the formation of
one-dimensional charge stripes.  Our experiments and analysis attribute the
weight loss around $\bar{M}$ to a FS crossing.  Indeed our
data of figure 1 (e) shows no evidence for a pseudogap in these
samples (the signature of a pseudogap in ARPES is that at $k=k_{F}$ the
leading edge of the
spectrum is depressed from $E_{F}$). Figure 1(e) shows that the midpoint of
our leading edge is fully up to or even past $E_{F}$. Our data also indicates
that the periodic arrangement of high intensity cusps
are due to the superstructure bands.  Mesot et al have also argued
that superstructure bands and not stripes are important for
understanding Saini's data\cite{18}.  However, in their picture the data is
analyzed starting from the standard hole-like FS and so they are not
capable of explaining the near-complete weight loss observed at $\bar{M}$.

Figs. 2 (h) and (i) show that the new behavior detailed here is robust as a
function of sample type.  These plots show $n(\vec k)$ and $E_{F}(\vec
k)$ along $\Gamma -\bar{M} - Z$ at hv's near
33 eV for a heavily
underdoped Bi2212 sample and for an overdoped
sample of the single-layer compound Bi2201. In both cases the data
indicates a main band crossing between $\Gamma$ and $\bar{M}$, indicating
that the FS
topology should also be electron-like for these samples (closed around
$\Gamma$) (the crossing is more washed out for the underdoped sample,
but appears to be qualitatively similar in other respects).  This indicates
that the new physics shown here is not
peculiar to one doping level or sample type.  Rather, we
argue that it is a product of the new photon energy range used
for our measurements.

Changing the photon energy in an ARPES experiment can have a number
of effects.  For a fixed $k_{\|}$, the most obvious effect is that
$k_{\perp}$ will change since the magnitude of the total momentum must
change. In this way, variations
in the electronic structure vs. $k_{\perp}$ may be mapped out, i.e. we can
map out the full three-dimensional electronic structure.  We have
taken data at many more photon energies (not shown here) to check whether
the Fermi
surface topology oscillates as a function of $k_{\perp}$.  We did not observe
any clear effects with a periodicity in $k_{\perp}$ of $2\pi/c$ where $c$
is the c-axis lattice constant.  Thus we conclude that the
differences in the data as a function of photon energy are not
naturally linkable to a coherent three-dimensionality of the band
structure.  This is consistent with the huge in-plane vs. out-of-plane
transport anisotropy of these materials.

Figure 4a shows the generally accepted $E$ vs. $\vec k$ relation for
near-optimal BSCCO, with a key feature being the large $\vec
k$-space region near $\bar{M}$ which has flat bands just below
$E_{F}$. Our figure 2g indicates that the crossing along $\bar{M}-Y$
is not robust, so we conclude that figure 4a is not a good
representation of the physics.  Rather, we propose the scenario in
figures 4b and 4c. The main band dispersion shown by the red lines is
as we have measured at $33 eV$, and is electron-like with a small
saddle-point at $\bar{M}$ above $E_{F}$.  Additionally, there is a
large region of flat bands at $\bar{M}$ which is observed at $22 eV$
but not $33 eV$. At 22 eV these states act to mask the true crossing
behavior of the bands observed at $33eV$, giving the impression of a
hole-like FS.  An important goal for future studies will be to
elucidate the origin of these additional states, including whether or
not they are intrinsic to the basic electronic structure of BSCCO.
Possible origins include quantum confinement due to stripe
formation\cite{19},
a contribution from the BiO states\cite{20,21},
indirect transitions \cite{22}, photoelectron diffraction effects
\cite{23}, final state (i.e. matrix element) effects \cite{22}, or
possibly a new type of correlated electron state due for instance to
magnetism.

Considering that electron transport should be dominated by the
dispersive states we have observed at $33eV$, we are left with the puzzling
result that the FS topology
looks electron-like, while Hall effect measurements indicate that the
carriers should be
hole-like\cite{24}. A similar disagreement has been
reported for the n-type superconductor $Nd_{2-x}Ce_{x}CuO_{4}$ - Hall
effect measurements have
indicated an electron-like FS \cite{25} while ARPES results have indicated a
hole-like FS\cite{26}.  These results indicate a non-simple relationship
between the Hall resistance and the electronic structure in these
materials.  This again highlights the importance of physics beyond
the band structure (i.e. correlation effects) in the cuprates.

In conclusion, by invoking a more complete set of FS crossing
criteria and by going to a non-traditional photon energy range we have
discovered an electron-like FS topology and the absence of flat bands at
$E_{F}$ in
the BSCCO family of hole-doped superconductors.  The
data is robust as a function of doping and is clearer than the
previous data which was interpreted as indicating a hole-like topology.
The main dissimilarity in the data sets occurs near $\bar{M}$, which is a
critical location in the Brillouin zone - it is where both the
superconducting gap and normal state
pseudogap were found to reach maximum amplitude\cite{6,7,8,9,10}.
Clearly this calls for more experimental and theoretical works considering
this new FS topology as
well as the effects of varying photon energy.

We acknowledge support from the Office of Naval Research's Young
Investigator Program.  The SRC is supported by the National Science
Foundation, and SSRL is operated by the Department of Energy, Office of Basic
Energy Sciences. We thank Aharon Kapitulnik for the use of one Bi2212 sample.
 
\end{multicols}
\vspace*{-0.2in}

%% Fig.1
\begin{figure}
 \caption{Normal-state ARPES from Bi2212. (a,b)  One quarter of the
 Brillouin zone showing (a) the hole-like FS extrapolated by Ding et al.
 [11] and (b) the electron-like FS determined here. Dark lines are
 the main FS pieces while the light lines represent superstructure-derived
 FS replicas.  (c,d) Raw data from Ding taken at $h\nu=19eV$ along directions
 indicated in panel (a).  (e) A portion of our new data taken at
 $h\nu=33eV$, with each $\vec k$-space point listed as a fraction $(x\pi)$ of
 the $\Gamma -\bar{M}$ distance.}
\end{figure}
 \vspace*{-0.2in}

% Fig.2
\begin{figure}
  \caption{(a-d): Schematics of the expected behavior of
  $n(\vec k)$ and $E_{F}(\vec k)$ for a band crossing the FS.
  (a) Non-interacting system at zero temperature. (b) Interacting
  Fermi Liquid with a reduced quasiparticle weight Z.  (c) Same as (b) but
with finite
  energy and momentum resolution.  (d) Same as (c) but including polarization
  and matrix element effects. Panels (e)-(i) are $n(\vec k)$ (red solid
  line) and
  $E_{F}(\vec k)$ (blue dashed line) at different photon energies from
BSCCO samples
  with varying number of $CuO_{2}$ layers and doping levels (indicated
  on panels). Green dashed lines indicate a main-band FS crossing.
  Superstructure-derived crossings are labeled s.s..}
\end{figure}
 \vspace*{-0.2in}

% Fig.3
\begin{figure}
  \caption{FS crossing points (white) from $33eV$ data on slightly overdoped
  Bi2212.  The thick black lines through the points
  indicate the main FS and the thin black lines indicate the
superstructure-derived
  FS.  Our data is overlayed on a color-scale plot of ARPES intensity
  at $E_{F}$ from an optimally doped Bi2212 sample, as measured at
  $33eV$ by Saini et al [12].  The highest intensity regions are yellow
  and the lowest intensity regions are black.}
\end{figure}

% Fig.4
\begin{figure}
  \caption{Schematic diagrams of the $E$ vs. $\vec k$ relations for
  near-optimal BSCCO.  (a) The old picture of the extended saddle point
  below $E_{F}$. (b,c) New dispersion relation showing a non-extended
  saddle point above $E_{F}$. We hypothesize an additional set of
  non-dispersive states below $E_{F}$ which are visible at $22eV$ but
  not $33eV$, and which sometimes mask the true crossing behavior.}
  \end{figure}

\end{document}